\documentclass[journal,comsoc]{IEEEtran}
\usepackage[T1]{fontenc}
\usepackage{enumitem}
\usepackage{multirow}
\usepackage{cite}
\ifCLASSINFOpdf
  \usepackage[pdftex]{graphicx}
  \graphicspath{{figures/}}
  \DeclareGraphicsExtensions{.pdf,.jpeg,.png}
  \else
  \usepackage[dvips]{graphicx}
  \graphicspath{{figures/}}
  \DeclareGraphicsExtensions{.eps}
  \fi
\usepackage{amsmath}
\usepackage[cmintegrals]{newtxmath}
\usepackage{array}
\usepackage[]{hyperref}
\hypersetup{
}
\usepackage{xcolor}

\usepackage{booktabs, threeparttable}
\usepackage{array}
\newcolumntype{L}[1]{>{\raggedright\arraybackslash}p{#1}}

\begin{document}
\title{Examining Machine Learning for 5G and Beyond through an Adversarial Lens}

\author{
\IEEEauthorblockN{Muhammad Usama$^1$, Rupendra Nath Mitra$^2$, Inaam Ilahi$^1$, Junaid Qadir$^1$, and Mahesh K. Marina$^2$\\}
\IEEEauthorblockA{Information Technology University (ITU), Pakistan$^1$, \\ The University of Edinburgh, UK$^2$}
}

\maketitle


\begin{abstract}
Spurred by the recent advances in deep learning to harness rich information hidden in large volumes of data and to tackle problems that are hard to model/solve (e.g., resource allocation problems), there is currently tremendous excitement in the mobile networks domain around the transformative potential of data-driven AI/ML based network automation, control and analytics for 5G and beyond. In this article, we present a cautionary perspective on the use of AI/ML in the 5G context by highlighting the adversarial dimension spanning multiple types of ML (supervised/unsupervised/RL) and support this through three case studies. We also discuss approaches to mitigate this adversarial ML risk, offer guidelines for evaluating the robustness of ML models, and call attention to issues surrounding ML oriented research in 5G more generally. 


\end{abstract}

\section{Introduction}

A considerable amount of industry and academic research and development endeavors are currently paving the way toward 5G and Beyond 5G (B5G) networks. 5G networks, unlike its 4G counterpart, are foreseen to be the underpinning infrastructure for a diverse set of future cellular services well beyond mobile broadband to span multiple vertical industries. To flexibly and cost-effectively support diverse use-cases and to enable complex network functions at scale, 5G network design espouses several innovations and technologies such as artificial intelligence (AI) along with software-defined networking (SDN), network function virtualization (NFV), multi-access edge computing (MEC), and cloud-native architecture that are new to the domain of mobile telecommunications. 

Technical developments toward 5G and B5G of mobile networks are quickly embracing a variety of deep learning (DL) algorithms as a de facto approach to help tackle the growing complexities of the network problems. However, the well-known vulnerability of the DL models to the adversarial machine learning (ML) attacks can significantly contribute to broadening the overall attack surface for 5G and beyond networks. This observation motivates us to deviate from the on-going trend of developing a newer ML model to address a 5G network problem and, instead, examine the robustness of the existing ML models in relation to the 5G networks under adversarial ML attacks. In particular, we focus on representative use cases for deep neural network (DNN)-driven supervised learning (SL), unsupervised learning (UL), and reinforcement learning (RL) techniques in the 5G setting and highlight their brittleness when subject to adversarial ML attacks.     

Through this article, we would like to draw the attention of the research community and all stakeholders of 5G and beyond mobile networks to seriously consider the security risks that emerge from the rapid unvetted adoption of DL algorithms across the wide spectrum of network operations, control, and automation, and urge to make robustness of the ML models a criterion before they are integrated into deployed systems. Overall, we make the following two contributions. 

\begin{enumerate}
    \item We highlight that despite the well-known vulnerability of DL models to adversarial ML attacks, there is dearth of critical scrutiny on the impact of the wide-scale adoption of ML techniques on security attack surface of 5G and B5G networks.
    \item We bridge the aforementioned gap through a vulnerability study of the DL models in all its major incarnations (SL, UL, and Deep RL) from an adversarial ML perspective in the context of 5G and B5G networks. 
\end{enumerate}


\section{Background}

\subsection{Primer on 5G Architecture}

\begin{figure*}[ht]
    \centering    {\includegraphics[width= 1\textwidth]{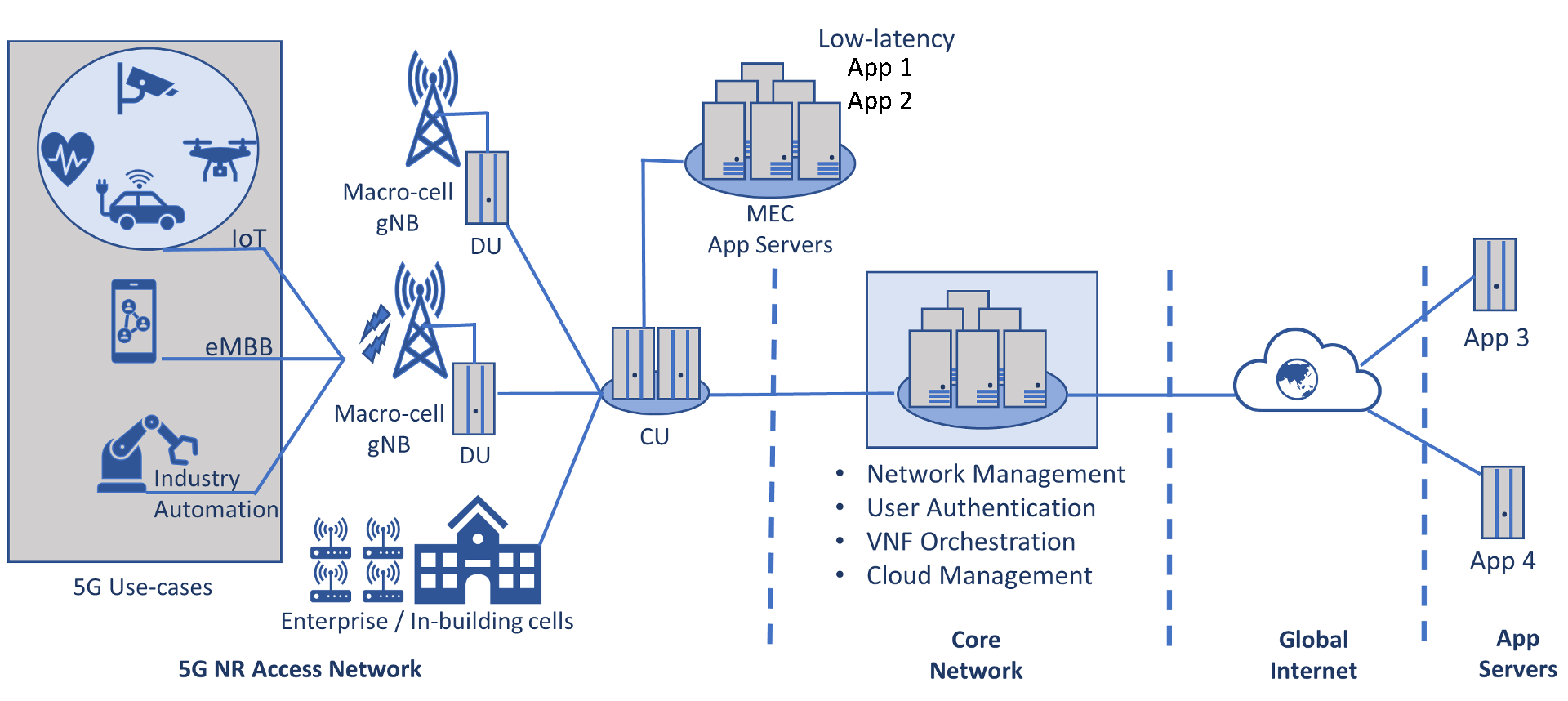}}
    \caption{A schematic diagram of 5G network architecture illustrating the disaggregated RAN architecture with distributed unit (DU) and centralized unit (CU) components; the MEC for improved latency; and the cloud-native core network and system orchestration components.}
    \label{fig:2}
\end{figure*}

A schematic diagram of the 5G network architecture is depicted in Figure 1. Apart from the user equipment (UE), the 5G system features a cloud-native core network, a flexible and disaggregated radio access network (RAN), and a provision for multi-access edge (MEC) cloud for reduced latency. The gNodeB (gNB) of the RAN comprises of split-able access nodes, distributed and centralized units (DU and CU), to efficiently handle evolved network requirements. The gNB connects to the MEC to significantly reduce the network latency for selected applications by availing edge server computing at the MEC cloud which is close to the radio service cells. For instance, to cater to the ultra-reliable low-latency communication (URLLC) use-case of industry automation, the RAN radio unit along with the DU, CU, and the MEC can be installed onsite. Thus, 5G network architecture enables applications to be deployed remotely (App 3 and App 4) or near the edge (App 1 and App 2) where low latency is a requirement. The provision of MEC also reduces the aggregated traffic load on the transport networks that is responsible for connecting RAN to the core network. The 5G core network (5G-CN) is a cloud-native network that stores subscriber databases and hosts essential virtualized network functions for network operations and management. Although, the network management and control functions are shown to be co-located with the core in the figure, they can be flexibly deployed at the edge as needed.

\subsection{ML in 5G and B5G Networks}

 A wide spectrum of DL algorithms are being developed for the broad context of wireless communications and 5G networking to deal with problems that are either hard to solve or hard to model \cite{shafin2020artificial}. For instance, optimal physical network resource allocation for NFV is an NP-hard problem and so require exponential computational power with increasing system size~\cite{haider2009challenges}. Deep RL (DRL)-based solutions are proposed to efficiently address resource allocation problems (e.g., \cite{foukas2019iris}). Network channel estimation for efficient beamforming is a problem that is hard to model and deep neural network (DNN)-based SL solution is well-accepted to tackle it \cite{huang2018deep}. Moreover, in certain use-cases, conventional expert systems become inappropriate due to real-world constraints, such as limited availability of power, where AI can perform effectively. For instance, deep autoencoder based systems can replace the power-hungry RF chain hardware with small embedded sensor systems enabling them to sustain longer on onboard power supplies \cite{Xmisc}. DL algorithms generally outperform the conventional approaches in solving mobile network prediction problems such as physical layer channel prediction by SL, signal detection problems such as recovering transmitted signals from noisy received signals by UL, and optimization problems like resource allocation by RL. 

\begin{figure*}[ht]
    \centering    {\includegraphics[width= 1\textwidth]{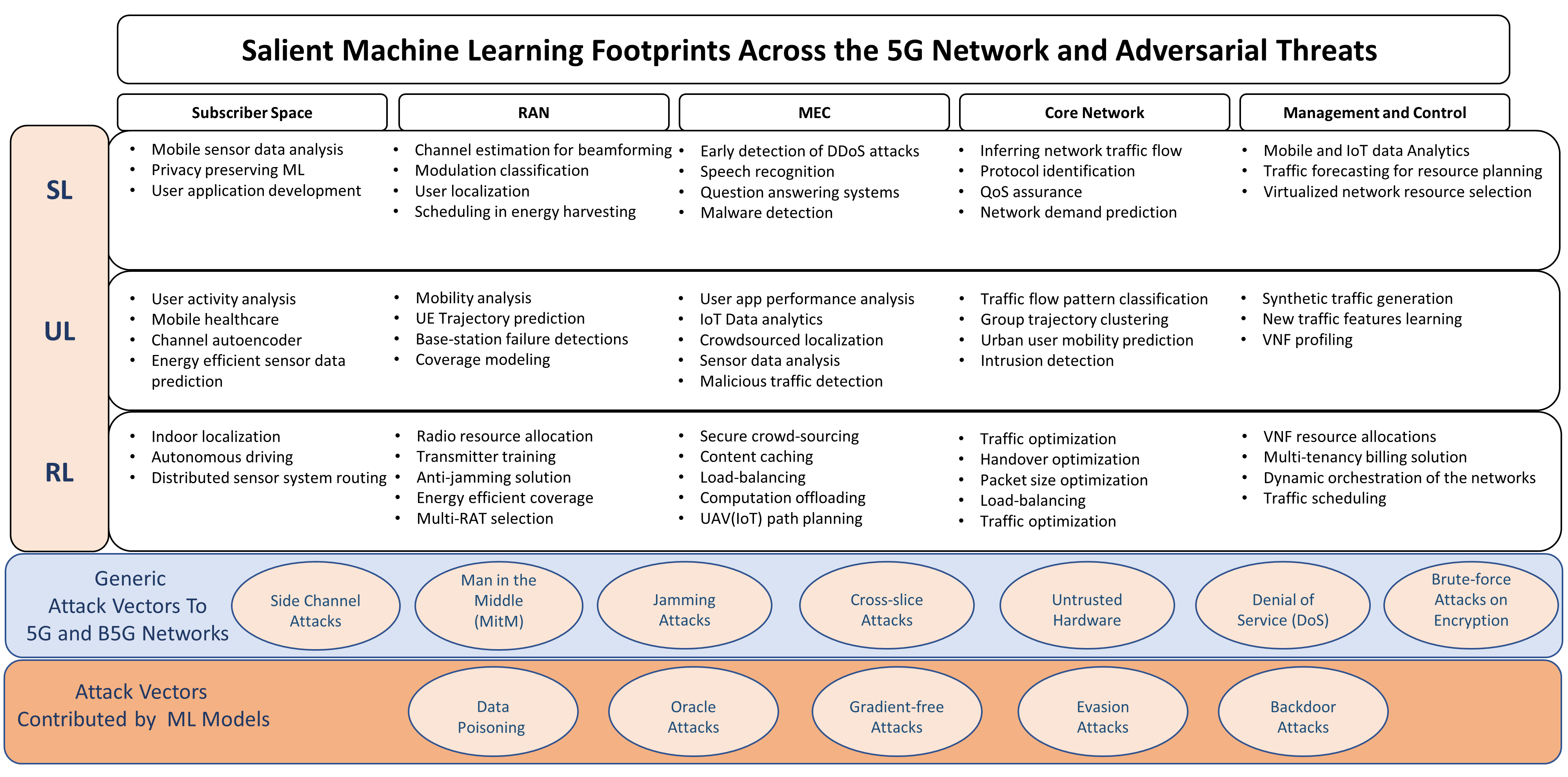}}
      \caption{Applicability of ML across the 5G network architecture and a depiction of how ML models contribute to significantly enhance the attack vectors beyond the traditional security risks through new adversarial ML risks \cite{ahmad2019security} \cite{biggio2018wild}}
    \label{fig:2}
\end{figure*}

\section{Widened Attack Surface in ML-Driven 5G and B5G Networks}


The security of the 5G networks is well explored (e.g., \cite{ahmad2019security}), but little attention has focused on the security of 5G and B5G networks in the face of the adversarial ML threat \cite{sagduyu2020wireless}. In this section, we briefly introduce the adversarial ML in general and subsequently outline the adversarial ML risks in 5G and B5G networks. 



\subsection{Overview of Security Attacks on ML} \label{sub:tax}
The vulnerability of the ML algorithms, especially the DL models, to the adversarial attacks is now well-established, where adversarial inputs are small carefully-crafted perturbations in the test data built for fooling the underlying ML model into making wrong decisions. An adversary can often successfully target an ML model with no knowledge of the model (\textit{black-box attack}), or some knowledge (\textit{grey-box attack}), or full knowledge (\textit{white-box attack}) of the target model. An adversary can attack the model during its training phase and in its testing phase as well. The training phase attacks are known as \textit{``poisoning attacks"} and the test time attacks are known as \textit{``evasion attacks"}. Evasion attacks are commonly known as adversarial attacks in the literature \cite{9003212}.

 More formally, an \textit{adversarial example} $x^*$ is crafted by adding a small indistinguishable perturbation $\delta$ to the test example $x$ of a trained ML classifier $f(.)$ where $\delta$ is approximated by the nonlinear optimization problem provided in equation \ref{eq1}, where $t$ is the class label.
\begin{equation}\label{eq1}
    x^* = x + \arg \underset{\delta{_x}}{\text{min}} \{\|\delta\|: f(x + \delta) = t\} 
\end{equation}

In 2013, Szegedy et al. \cite{szegedy2013intriguing} observed the discontinuity in the DNN's input-output mapping and reported that DNN is not resilient to the small changes in the input. Following on this discontinuity Goodfellow et al. \cite{goodfellow6572explaining} propose a gradient-based optimization method for crafting adversarial examples. This technique is known as \textit{fast gradient sign method} (FGSM). Papernot et al. \cite{papernot2016limitations} craft adversarial perturbation using a saliency map-based approach on the forward derivatives of DNN. This approach is known as \textit{Jacobian saliency map based attack} (JSMA). Carlini et al. \cite{carlini2017towards} crafted three different adversarial attacks using three different distance matrices ($L_1$, $L_2$, and $L_\infty$). More details about adversarial ML attacks are described in \cite{9003212, biggio2018wild}. 

\subsection{Added Threat from Adversarial ML for 5G and Beyond}

5G and B5G networks embrace ML-driven solutions for improved accuracy and smarter network operations at scale. Figure \ref{fig:2} illustrates network problems from different network segments of 5G, namely user devices, RAN, MEC, core networks, and the network management and control layer that have recently attracted ML-based solutions from all the three categories of ML. However, in light of the above discussion in \S \ref{sub:tax}, the DL-powered ML models gaining popularity for 5G and B5G networks are vulnerable to the adversarial attacks thereby further aggravating the security risks of future generations of mobile networks. 

In our attempt to show the feasibility of adversarial ML attacks on 5G systems we take three well-known ML models---one from each of the three ML families of algorithms (UL, SL, and DRL)---from wireless physical layer operations relevant to 5G and B5G context and show the vulnerability that naive use of ML brings to future mobile networks. We choose all the three ML models for our case studies from the physical layer network operations because of the maturity of ML-research in the context of AI-driven 5G networking and the availability of open-sourced ML models backed up with accessible data-sets\footnote{\url{https://mlc.committees.comsoc.org/research-library/}}.


\section{Highlighting Adversarial ML Risk for 5G and Beyond: Three Case Studies} \label{sec:3}

In this section, we critically evaluate to exemplify the security threats posed by DL models in the three canonical ML families of algorithms (UL, SL, and RL) in the 5G and B5G networking and present our work as three different case-studies.

\subsection{Attacking \textbf{Supervised ML-based} 5G Applications}\label{3a}
Automatic modulation classification is a critical task for intelligent radio receivers where the signal amplitude, carrier frequency, phase offsets, and distribution of noise power are unknown variables to the receivers subjected to real-world frequency-selective time-varying channels perturbed by multipath fading and shadowing. The conventional maximum-likelihood and feature-based solutions are often infeasible due to the high computational overhead and domain-expertize that is involved.  To make modulation classifiers more common in modern 5G and B5G networked devices, current approaches deploy DL to build an end-to-end modulation classification systems capable of automatic extraction of signal features in the wild \cite{meng2018automatic}.

\begin{figure}[h]
  \centering
  \includegraphics[width=\linewidth]{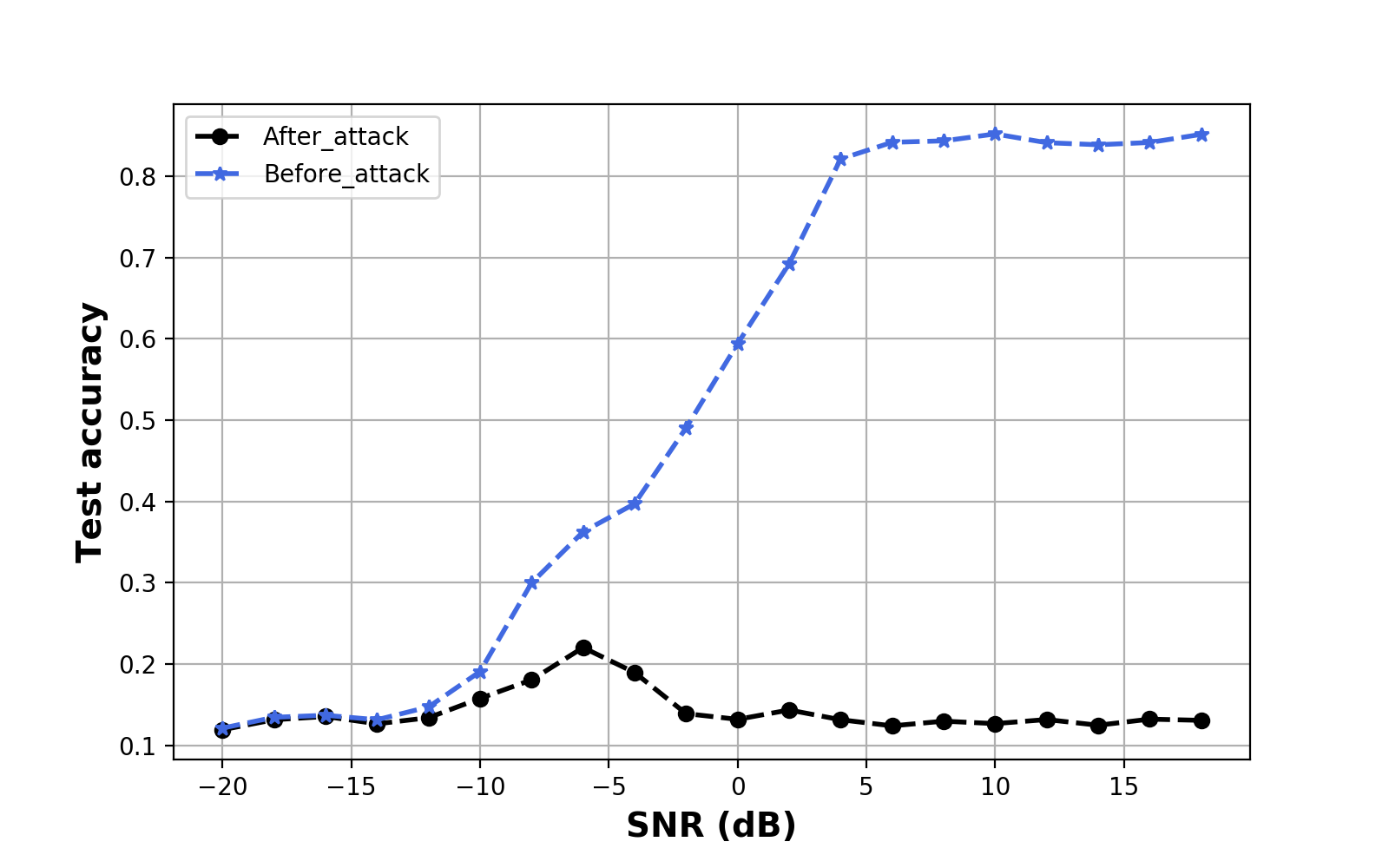}
  \caption{Accuracy of the CNN-based automatic modulation classifier before and after the adversarial ML attack. \textit{A clear drop in the accuracy of the classifier with the increasing SNR indicates the success of the adversary in compromising the integrity of the modulation classifier that is seen as viable in the 5G and B5G networks}.}
  \label{fig:mod-class}
\end{figure}

 We pick a convolutional neural network (CNN)-driven SL-based modulation classification model in this case study to illustrate the added dimension of vulnerability introduced in the networks by it. We use the well-known GNU radio ML RML2016.10a dataset that consists of 220000 input examples of 11 digital and analog modulation schemes (AM-DSB, AM-SSB, WBFM, PAM4, BPSK, QPSK, 8PSK, QAM16, QAM64, CPFSK, and GFSK) on the signal to noise ratio (SNR) ranging from -20 dB to 18dB \cite{o2016radio}.  However, we exclude the analog modulation schemes from our study and consider only the eight digital modulations from the data set because from 2G onward all mobile wireless standards are strictly digital communications. Figure \ref{fig:mod-class} depicts the classification performance of the CNN model in the multi-class modulation classification for the signals between -20dB to 18dB of SNR. 
 
To show the feasibility of an adversarial ML attack on the CNN-based modulation classifier we make the following assumptions:
\begin{itemize}
    \item We consider the \textit{white-box} attack model where we assume that the adversary has a complete knowledge about the deployed modulation classifier.
    \item \textit{Goal of the adversary} is to compromise the integrity of the CNN classifier leading to a significant decay in the classification accuracy which is the measure of the \textit{success of the adversary}.
\end{itemize}

To craft the adversarial examples to fool the CNN classifier, we use the Carlini \& Wagner (C\&W) attack \cite{carlini2017towards} for each modulation class by minimizing the $L{_2}$ norm on the perturbation $\delta$, such that when the perturbation $\delta$ is added to the input $x$ and sent to the CNN-based modulation classifier $C$ it misclassifies the input $x$. More details on the C\&W attack are available in \cite{carlini2017towards}. The performance of the CNN-based modulation classifier before and during the adversarial attack is depicted in Figure \ref{fig:mod-class}. A distinct drop in the accuracy of the modulation classification after the adversarial attacks indicates the brittleness of deep supervised ML in 5G and B5G applications. Moreover, our results show that the approbation of unsafe DL models in the physical layer operations of the 5G and B5G networks can make the air-interface of the future networks vulnerable to adversarial ML attacks.


\begin{figure*}[ht]
    \centering    {\includegraphics[width= 1\textwidth]{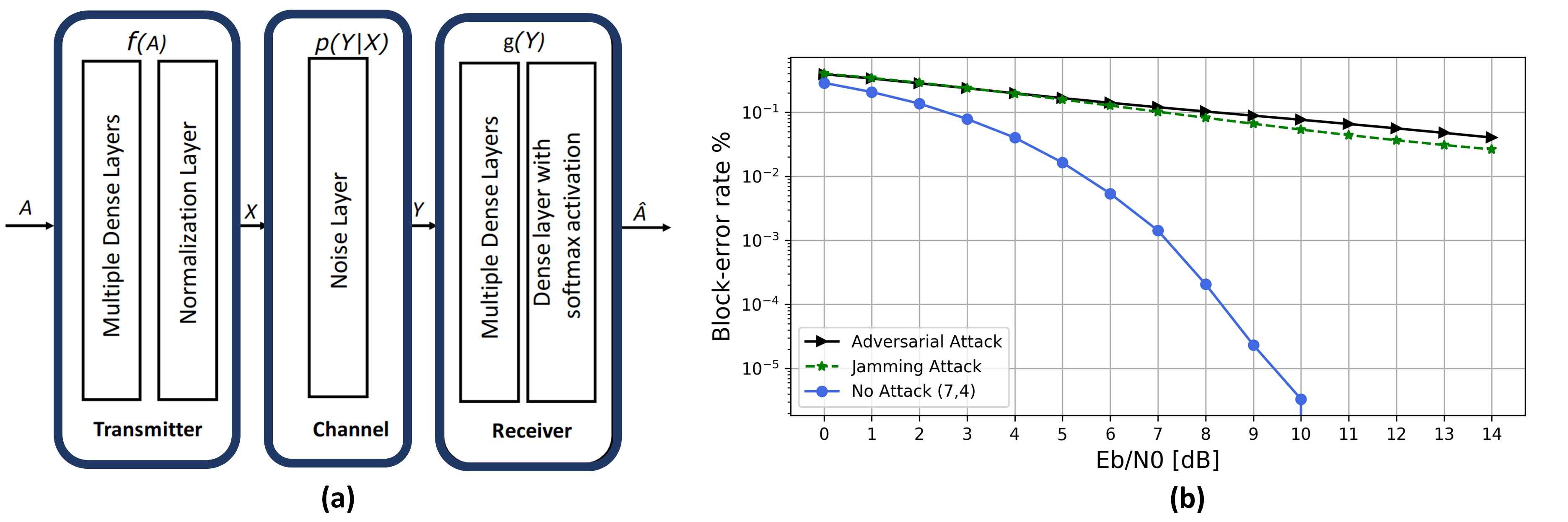}}
    \caption{(a) Architecture of channel autoencoder for 5G and future networks proposed in \cite{o2017introduction}; (b) Performance of the channel autoencoder before and under the adversarial ML attack and traditional jamming attack. \textit{The Block Error Rate (BLER) versus $E{_b}/N{_0}$ curves indicates that adversarial ML attack does not only deteriorate the model's performance but often outperformed the known jamming attack}.}
    \label{fig:4}
\end{figure*}

\subsection{Attacking \textbf{Unsupervised ML-based} 5G Applications}
\label{case-study 2}

In 2016, O'Shea et al. proposed the idea of channel autoencoders which is an abstraction of how an end-to-end radio communication module functions in real-world wireless systems \cite{o2016learning}. The deep autoencoder-based communication model gains rapid popularity and is seen as a viable alternative to the dedicated radio hardware in the future 5G and beyond networks \cite{Xmisc}. Figure \ref{fig:4}(a) depicts the conceptual design of the channel autoencoder that we choose as a deep UL model for this case study. We assume the model is subjected to an additive white Gaussian noise (AWGN) channel and apply the parameter-configurations provided in \cite{o2017introduction}. To perform the adversarial ML attack on the channel autoencoder we consider the following threat model and compare the performance of the model with and without attack.

\begin{itemize}
	\item We assume a \textit{white-box} setting, where the adversary has complete knowledge of the deployed ML model. We further assume that the autoencoder learns a broadcast channel. The proposed adversarial attack on channel autoencoder can be converted into a black-box adversarial attack, where the adversary has zero knowledge of the target ML model, by following the surrogate model approach provided in \cite{8766505}.
	\item The \textit{goal of the adversary} is to compromise the integrity of channel autoencoder and the \textit{success of the adversary} is measured by the elevated block error rate (BLER) with improving SNR per bit ($E_b/N{_0}$).
\end{itemize}

We take the following two-step data-independent approach to craft adversarial examples for the channel autoencoder:

\begin{enumerate}
    \item Sample the Gaussian distribution randomly (we sampled Gaussian distribution because the channel is AWGN)  and use it as an initial adversarial perturbation $\delta$;
    \item Maximize the mean activations of the decoder model when the input of the decoder is the perturbation $\delta$. 
\end{enumerate}

This produces maximal spurious activations at each decoder layer and results in the loss of the integrity of the channel autoencoder. Figure \ref{fig:4}(b) shows the performance of the model before and under the adversarial attack. Moreover, the figure suggests the adversarial ML attack is often outperforms the traditional jamming attacks.

Since the idea of channel autoencoder in a wireless device is to model the on-board communication system as an end-to-end optimizable operation, the adversarial ML attacks on channel autoencoder show that the application of unsupervised ML in the 5G mobile networks increases its vulnerability to adversarial examples. Hence, we argue that deep UL-based 5G networked systems and applications need to be revisited for their robustness before being integrated into the 5G IoT, and related systems.


\begin{figure*}[ht]
    \centering    {\includegraphics[width= 1\textwidth]{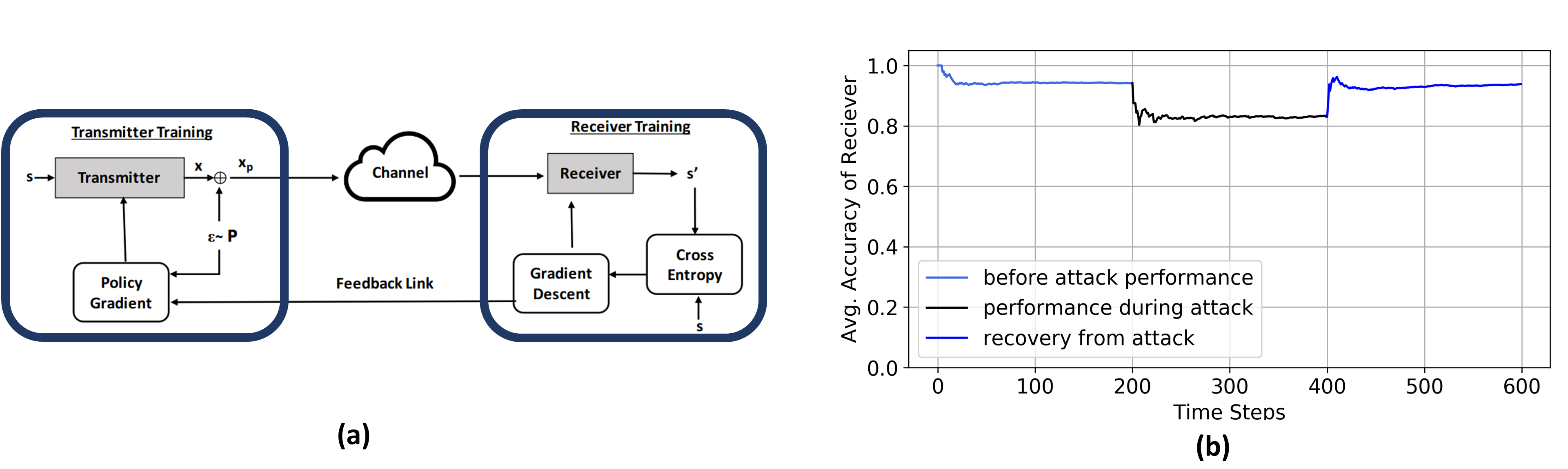}}
    \caption{(a) Architecture of DRL-based channel autoencoder with noisy feedback for 5G and future networks proposed in \cite{goutay2019deep}; (b) Performance of DRL autoencoder with noisy feedback before, during, and after the adversarial ML attack. \textit{A clear drop in the performance of the receiver during the attack indicates the success of the adversary in compromising the DRL autoencoder-based end-to-end communication system in future wireless networks}.}
    \label{fig:5}
\end{figure*}

\subsection{Attacking \textbf{DRL-based} 5G Applications}

In the final case study, we have performed the adversarial ML attacks on an end-to-end DRL autoencoder with a noisy channel feedback system \cite{goutay2019deep}. Before providing the details of the adversarial ML attack on the DRL-based communication system, we briefly discuss the DRL-based channel autoencoder. Goutay et al. \cite{goutay2019deep} take the same architecture we consider in the previous case study \S \ref{case-study 2} and add a noisy feedback mechanism to it, as shown in Figure \ref{fig:5}(a). The end-to-end training procedure involves:

\begin{enumerate}
    \item The RL-based transmitter training by a policy gradient theorem \cite{sutton2000policy} to ensure that the intelligent transmitter learns from the noisy feedback after a round of communication.
    \item SL model-based receiver training to train the receiver as a classifier.
\end{enumerate}

Both transmitter (encoder) and receiver (decoder) are implemented as separate parametric functions (differentiable DNN architectures) that can be optimized jointly. The communication channel is modeled as a stochastic system that provides a conditionally distributed relation between the encoder and the decoder of the channel autoencoder. More details on the design and training procedure are available in \cite{goutay2019deep}. The considered threat model for this case study is given as:



\begin{itemize}
	\item We choose a \textit{black-box} settings where the adversary does not know the target model. We assume black-box settings because these are realistic as the attacker has no information about the deployed encoder and decoder model, and the adversary can only attack by adding perturbation in the broadcast channel. We also assume that the adversary can perform an adversarial ML attack for ``$n$''-time steps. 
	\item The \textit{goal of the adversary} is to compromise the performance of the DRL autoencoder with noisy feedback for a specific time interval. The \textit{success of the adversary} is measured by the degradation in the decoder's performance during the attack interval.  
\end{itemize}

To evaluate the robustness of DRL model, we exploit the transferability property of the adversarial examples, which states that adversarial examples compromising an ML model will compromise other ML models with high probability if the underlying data distribution is same between these two models. We opt for the surrogate model approach for performing the adversarial ML attack. So we transfer the adversarial examples crafted in case study (\S \ref{case-study 2}) and measure the average accuracy of the receiver. We run the DRL autoencoder with a noisy feedback system for 600-time steps (one time-step is equal to one communication round) and perform the adversarial attack between 200 to 400-time step window. We transfer 200 successful perturbations from the previous case study (\S \ref{case-study 2}) and report the performance for the DRL system. Figure \ref{fig:5}(b) provides the performance of the receiver (decoder) of the DRL autoencoder. It is evident that the performance of the receiver degrades from 95\% to nearly 80\% during the adversarial attack window. The recovery in the performance after the adversarial attack explicates that the end-to-end DRL autoencoder communication system can recover because of the noisy feedback and DRL adaptive learning behavior.

Our results, as presented in this section, confirm the feasibility of adversarial ML attacks on DL-based applications from all the three types of ML algorithms that are prevalent in the 5G network systems and highlight the additional threat landscape emerges due to the integration of vulnerable DL models to the 5G and B5G networks.


\section{Discussion}

In this section, we discuss the inadequacies of ML models to address the comprehensive technology needs of 5G networks and we recommend a set of improving measures toward robust ML models appropriate for the 5G and B5G approbation.

\subsection{Towards Robust ML-Driven 5G and Beyond Networks}

Robustness against adversarial ML attacks is a very challenging problem. To date, there does not exist a defense that ensures complete protection against adversarial ML attacks. In our previous works \cite{9003212, ilahi2020challenges}, we have performed an extensive survey of the adversarial ML literature on robustness against adversarial examples, and showed that nearly all defensive measures proposed in the literature can be divided into: 
\begin{enumerate}
\item modifying data approaches (adversarial training, feature squeezing, input masking, etc.); 
\item auxiliary model addition approaches (generative model addition, ensemble defenses, etc.); 
\item modifying model approaches (defensive distillation, model masking, gradient regularization, etc.).
\end{enumerate}


Our results in \S \ref{sec:3} indicate that ML-based 5G applications are very vulnerable to the adversarial ML attacks. There does not exist much work on the recommendations and guidelines for evaluating the robustness of ML in 5G applications. In the following, we have provided a few important evaluation guidelines for evaluating the ML-based 5G applications against adversarial ML attacks. These insights are extracted from the Carlini et al. \cite{carlini2019evaluating} and our previous works \cite{8884228, 9003212}. 
\begin{itemize}
	\item Many defenses are available in the literature against adversarial attacks but these defenses are limited by the design of the application. Using them without considering the threat model of ML-based 5G applications can create a false sense of security. So for ML-based 5G application threat models must clearly state the assumptions taken, type of the adversary, and the metrics used for evaluating the defense. 
    \item Always test the defense against the strongest known attack and use it as a baseline. Evaluating for an adaptive adversary is also necessary.
    \item Evaluate the defense procedure for gradient-based, gradient-free, and random noise-based attacks \footnote{\url{https://www.robust-ml.org/}}. 
    \item Clearly state the evaluation parameters (accuracy, recall, precision, F1 score, ROC, etc.) used in validating the defense and always look for a change in the false positive and false negative scores. 
    \item Evaluation of the defense mechanism against out-of-distribution examples and transferability-based adversarial attacks is very important. 
\end{itemize}
Although these recommendations and many others in \cite{8884228, 9003212, ilahi2020challenges, carlini2019evaluating} can help in designing a suitable defense against adversarial examples but this is still an open research problem in adversarial ML and ripe for investigation for ML-based 5G applications. 

\subsection{Beyond Vulnerability to Adversarial ML Attacks}

Apart from the vulnerability of the ML models to the adversarial ML attacks, we underline the following drawbacks that call into question the viability of ML-driven solutions to be integrated into the real-world 5G networks any time soon.

\vspace{1mm}
\subsubsection{Lack of real-world datasets}
The availability of large data from real-world sources is the fuel of the ML models. Especially the advancement of modern DL research critically depends on having easy access to a variety of data-sets in the research community. However, diversified real-world data-sets in telecommunication and mobile networking is not readily available due to privacy issues and stringent data-sharing policies adopted by the global telecom operators. Hence, a large amount of ML research in the telecom domain still depends on synthetic data which often falls short of truly representing real-world randomness and variations. Thus, current state-of-the-art ML models in telecommunication applications are oftentimes can not replace the domain-knowledge based expert systems currently in operation.

\vspace{1mm}
\subsubsection{Lack of explainability}
In ML studies, the accuracy of a model comes at the cost of explainability. The DL models are highly accurate in providing output but lack an explanation of why a particular output is achieved. Explanation of a decision taken often becomes a critical requirement for the 5G network, especially because many critical services such as transport signaling, connected vehicles, and URLLC depend on the 5G infrastructure.


\vspace{1mm}
\subsubsection{Lack of operational success of ML in real-world mobile networks}
A plethora of ML models exists in the literature of mobile networking, albeit there is a dearth of operational ML models in currently operational mobile networks. We perform attacks on the ML models running under the ideal environment, simulated in favorable lab conditions, and still, the victim models can not withstand the adversaries. In the real-world mobile network, the ML models need to be deployed and functional under unforeseen random environments leaving them more vulnerable to the cyber-attacks. Moreover, the computational overhead, requirement of hardware enhancement, and run-time delay introduced by the ML models become critical factors to bring operational success for ML in the real-world large-scale mobile networks like 5G.


\section{Conclusions}
Security and privacy are uncompromising necessities for a modern and future global networks standards such as 5G and Beyond 5G (B5G) and accordingly there is an interest in fortifying it to thwart attacks on it and withstand the rapidly evolving landscape of future security threats. This article specifically highlights the approbation of a large number of DL-driven solutions in 5G and B5G networking gives rise to security concerns that remain unattended by the 5G standardization bodies, such as the 3GPP.  We argue this is the right time for cross-disciplinary research endeavors considering ML and cybersecurity to gain momentum to enable secure and trusted future 5G and B5G mobile networks for all future stakeholders. We hope that our work will motivate further research toward a ``telecom-grade ML'' that is safe and trustworthy enough to be incorporated into 5G and beyond 5G networks, thereby power intelligent and robust mobile networks supporting diverse services including mission-critical systems.


\makeatother

\bibliographystyle{unsrt}
\bibliography{ref}




\end{document}